# Plasmonic resonators for enhanced diamond NV-center single photon sources

Irfan Bulu*, Thomas Babinec, Birgit Hausmann, Jennifer T. Choy, and Marko Loncar

*School of Engineering and Applied Sciences, Harvard University, Cambridge, MA 02138*
*irfan@seas.harvard.edu*

**Abstract:** We propose a novel source of non-classical light consisting of plasmonic aperture with single-crystal diamond containing a single Nitrogen-Vacancy (NV) color center. Theoretical calculations of optimal structures show that these devices can simultaneously enhance optical pumping by a factor of 7, spontaneous emission rates by Fp ~ 50 (Purcell factor), and offer collection efficiencies up to 40%. These excitation and collection enhancements occur over a broad range of wavelengths (~30nm), and are independently tunable with device geometry, across the excitation (~530nm) and emission (~600-800nm) spectrum of the NV center. Implementing this system with top-down techniques in bulk diamond crystals will provide a scalable architecture for a myriad of diamond NV center applications.

**Introduction**

Diamond nitrogen vacancy color center (NV center) has recently emerged as a promising platform for realization of scalable solid-state quantum networks, sensitive magnetometry, and secure communication [1-7]. NV center, formed by a substitutional nitrogen atom and a proximal missing carbon atom in diamond lattice, has a sharp, zero phonon line (ZPL), emission around 637 nm with a broad phonon sideband that extends well into the NIR region of the spectrum [1]. The defect center has rather interesting spin properties. The spin states of the center exhibit long coherence times. In addition, the spin state can be initialized by non-resonant optical pumping and can be accessed optically [3,8-10].

Nearly all applications of the diamond NV center rely on its efficient optical initialization and detection. Hence, their performance strongly depends on the ability to efficiently excite, generate, and collect photons from a single NV center. For instance the sensitivity of a diamond NV center magnetometer can be directly linked to the photon emission rates and collection efficiencies [4]. Therefore, it is of great interest to improve the efficiency of photon in- and out-coupling using optical nanostructures. For example, waveguiding in a diamond nanowire with embedded NV fabricated using top-down processes or a diamond solid immersion lens allows for efficient excitation and collection in a scalable system, but without the benefit of engineered radiative recombination that is provided by an optical cavity [11,12]. In addition, plasmonic nanowires and optical antennas have already been demonstrated, and enhancement of the spontaneous emission rate of NV was verified [13,14]. However, previously explored plasmonic configurations required complex alignment and positioning techniques and are challenging to scale.

In this work, we propose a novel device platform consisting of diamond-plasmon apertures that allow for efficient excitation, increased photon emission rates, and enhanced collection of emitted photons relative to a bulk diamond sample. In addition, recent advances in top-down manufacturing of quantum optical nanostructures in bulk diamond will allow chip-scale fabrication of the proposed devices with realistic modifications to existing recipes [15-17]. We compare four different geometries, Fig.1, with varying degrees of performance and fabrication complexity. We show, theoretically, that in an optimized geometry pumping can be improved by a factor of 7, the spontaneous emission rates can be enhanced up to 50 times over NV in a bulk crystal, and collection efficiencies can be as high as 40% (more than ten-fold increase over the bulk). Compared to previously demonstrated plasmonic structures for quantum-diamond applications, our plasmonic apertures provide a top-down scalable architecture with enhanced performance.

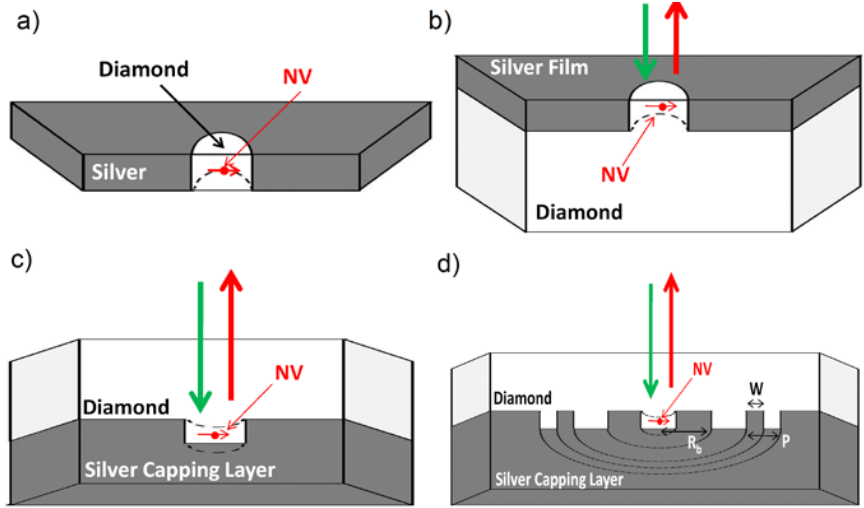

Fig. 1. 3-dimensional cross section of the plasmonic resonators that we consider in this work. The structures consist of circular diamond nanowires containing an NV center. The diamond nanowire is radially surrounded by silver. (a) The structure is surrounded by air on top and bottom. (b) A cylindrical diamond nanowire of finite height surrounded radially by silver. The structure is surrounded by air on top and by diamond on bottom. NV center is excited from the top side and photons are also collected from the top side. (c) The plasmonic resonator is on the bottom side of the diamond crystal. The plasmonic resonator is completely embedded in a thick layer of silver on the bottom side. The collection and excitation directions are shown by the red and green arrows, respectively. (d) Similar to the structure shown in Fig. 1(c). In this case, the plasmonic resonator is surrounded by a single silver ring or multiple silver rings. This configuration improves the collection efficiency.

**Results and Discussion**

First, we briefly discuss the properties of the guided modes of plasmonic apertures, formed by surrounding infinitely long diamond nanowires/ fibers, with circular cross-section, with metal [18,19]. A plasmonic aperture supports guided modes similar to those of dielectric waveguides, but has distinctly different confinement properties: the confinement factor of a dielectric waveguide *decreases* as the dimensions of the waveguide are decreased beyond a certain value, while the confinement of a plasmonic aperture *increases* as the dimensions are decreased. The confinement properties of a circular dielectric waveguide and a circular plasmonic aperture can be compared with the effective mode area, which is defined in a similar way to the more familiar term effective mode volume [20,21].

$$A_{\mathrm{mode}}(x_0, y_0) = \frac{\iint \mathrm{Re}\frac{\partial(\omega\varepsilon)}{\partial\omega}|E|^2\,dxdy}{\mathrm{Re}\frac{\partial(\omega\varepsilon(x_0,y_0))}{\partial\omega}|E(x_0,y_0)|^2} \tag{0.1}$$

We show the mode area for the HE11 mode, the fundamental mode, at 637 nm in Fig. 2 (a). Note that the mode area for the plasmonic aperture decreases as the aperture radius is decreased and it is as small as 0.04 $(\lambda/n)^2$ when the radius is 45 nm. We also show the mode area as a function of radius for a diamond nanowire for comparison. The mode for the diamond nanowire is significantly delocalized for radii smaller than 60 nm. Electric energy density for the plasmonic aperture is plotted in the inset of Fig. 2(a). Energy density is maximum in the center of the dielectric core and it is fairly uniform within the dielectric core. As a result, a radially polarized dipole is expected to have significant overlap with the aperture mode regardless of its radial position within the aperture. In addition, the effective mode index and propagation lengths are plotted as a function of aperture radius in Fig. 2(b) at a wavelength of 637 nm. The mode index is closer to 0 for smaller apertures and 2.4, index of the core material, for larger apertures. Propagation lengths are significantly short for very smal apertures.

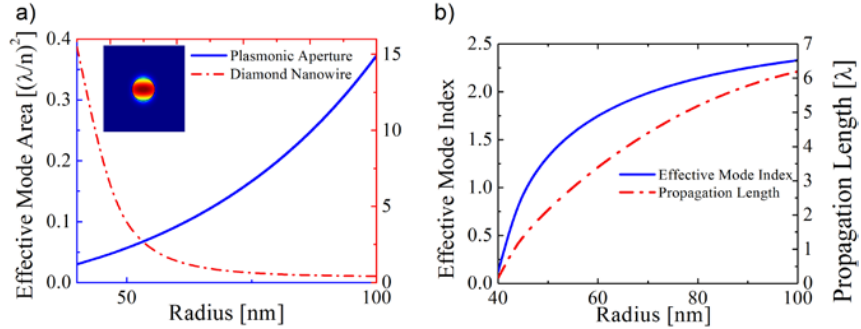

Fig. 2. (a) Mode area for the HE11 mode of a plasmonic aperture and a diamond nanowire at 637 nm. Inset: electric energy density distribution for the guided mode in the case of the aperture. (b) Effective mode index and propagation length in units of wavelength for plasmonic aperture.

By truncating the infinitely long metal-clad diamond fiber, we get the geometry shown in Fig. 1(a). The structure consists of a cylindrical diamond nanowire surrounded with silver on the sides, and air on the top and the bottom. We plot the spontaneous emission rate enhancement spectrum of a radially polarized dipole when it is placed at the center of the aperture as a function of radius and height in Figs. 3(a) and (b), respectively. We consider radially polarized dipoles throughout this paper. This polarization is appropriate for NV centers that can be found in [111] terminated diamond crystals. Permittivity values from Johnson&Christy [22] were used for silver and permittivity values from Palik [23] were used for diamond. SE rate enhancement factors were calculated by comparing the total power emitted from the dipole when it was placed in the aperture to the total power emission when it is in infinitely thick bulk diamond. We observe two resonances in the SE rate enhancement spectrum for this particular dipole position. The resonances are red shifted when the radius or the height of the aperture is increased. The response of the two resonances to a change in height is quite different. While the resonance at the shorter wavelengths red shifts almost linearly with an increase in height, the shift in the location of the second resonance is negligible for apertures taller than 250 nm. To explain this feature, we consider a simple Fabry-Perot model. The resonance condition is such that the total phase due to propagation and reflection is an integer multiple of $\pi$, $n_{eff} k_0 L + \Delta\Phi_R = m\pi$. We show $n_{eff} k_0 L + \Delta\Phi_R - m\pi$ in Fig. 3(c) for m=1-4 for an aperture with height 350 nm and radius 50 nm. The resonance condition is satisfied for m=1 at 696 nm and at 525 nm for m=4. The resonance wavelengths calculated from the full 3D solution of the Maxwell equations, Fig. 3(d), are in excellent agreement with the predictions from the Fabry-Perot model. The anomalous dependence of the first order Fabry-Perot mode to the height of the aperture can be attributed to the steep change in the total phase around 690 nm wavelength. Once the height of the aperture is such that the first order Fabry-Perot resonance is around 690 nm, an increase in height will only result in a very small change in the resonance wavelength of the first order mode. This property will be exploited in the following section in order to design doubly resonant apertures for diamond NV center applications. In Fig. 3(d) we show the SE rate enhancement as a function of the axial position of the dipole, for four different wavelengths corresponding to $1^{st}$, $2^{nd}$, $3^{rd}$ and $4^{th}$ resonance of the structure with height 350 nm and radius 50 nm. It can be seen that the shape of the SE-rate curves follow expected FP resonance profile. However, we note that the SE Rate enhancement factors do not vanish at the nodes of Fabry-Perot resonances (Fig. 3(d)), and it is as large as 2 for the $2^{nd}$ order mode. This is attributed to the direct coupling of the dipole to the radiation modes and surface plasmons that exist at the metal-air interface. The quality factor of the resonances can be determined from the decay rate of the stored energy or by fitting the resonance line shape to a Fano model [24,25]. The use of Fano model is necessary in order to take coupling into non-resonant channels into account. Our calculations showed that the difference between the two methods is less than 1%. We show the resonant frequencies and the corresponding quality factors as a function of the aperture height in Fig. 4(a). The resonant wavelengths scale linearly for short apertures and saturates for taller apertures. On the other hand, quality factors increase linearly with the aperture height. The loss limited quality factors are around 175. As a result, the quality factors are mostly limited by coupling in to radiative channels. This is desirable in order to be able to extract the emitted light before it gets dissipated in the form of heat due to metal losses.

Finally, SE rate enhancement can be estimated from $\dfrac{3}{2\pi^2}\dfrac{W_E}{W_E+W_M}\dfrac{Q}{V}\left(\dfrac{\lambda_0}{n}\right)^3$ where $W_E$ and $W_M$ are the stored electric and magnetic energies, respectively. Mode volume, V, has the same form as the effective mode area except the integration is over a large volume. $\dfrac{W_E}{W_E+W_M}$ term takes proper normalization of the modes of dispersive cavities into account. This point has also been recently discussed by Iwase [20]. The results are plotted in Fig. 4(b) along with the FDTD results for different aperture heights. There is a good agreement between the two results at larger aperture heights for which the quality factors are larger. The difference at shorter apertures is due to the non-resonant coupling into decay channels other than the cavity mode.

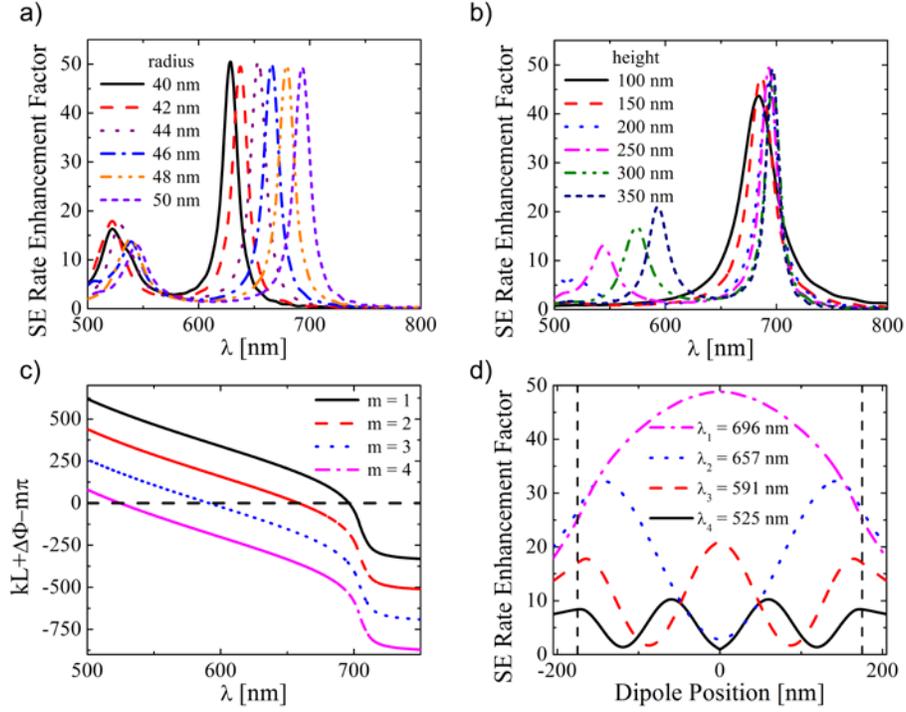

Fig. 3. SE Rate enhancement spectrum: (a) as a function of radius at fixed height, h = 250 nm, (b) as a function of height at fixed radius, r = 50 nm. The dipole is located at the center of the aperture. As a result, only odd order Fabry-Perot resonances are excited. (c) Fabry-Perot condition, i.e., the difference between $m\pi$ and the total phase, i.e., the phase due to propagation and reflection. r = 50 nm and h = 350nm. (d) SE Rate enhancement factor as a function of dipole position at resonance wavelengths for an aperture height of 250 nm and radius of 50 nm.

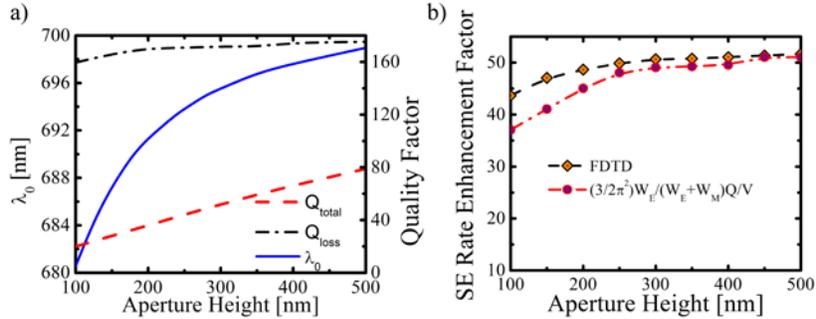

Fig. 4. (a) Resonant wavelengths and the corresponding quality factors as a function of aperture height for r = 50 nm. (b) SE rate enhancement factor from the modified Purcell formula and from FDTD as a function of height for r = 50 nm.

Encouraged by these results in the case of symmetric structure, we modified our geometry slightly to facilitate fabrication of the device. The modified structure, shown in Figure 1(b), can be fabricated using reactive ion etching followed by metal lift-off (using the same mask) [15]. In comparison, the symmetric structure shown in Fig. 1(a) requires use of very thin diamond slabs which are challenging to realize. The SE rate enhancement spectra, for modified geometry, for two different radii, r = 45 and 50 nm, are shown in Fig. 5(a). The height is 220 nm for both apertures. Since diamond NV center has a very broad spectrum, which consists of ZPL at 637nm wavelength, and phonon side-band with maximum emission around 680nm, we designed the first aperture to have a good overlap with the ZPL, and the second one with the maximum emission. In both cases, the dipole is 90 nm below the air-diamond interface. Using numerical modeling, we found that spontaneous emission rate is enhanced by more than 50 times in the first case (ZPL emission) and more than 30 times in the second case (~680 nm). The first resonator is designed such that there is also a resonance around 532 nm. We note that this is the common choice of excitation wavelength for diamond NV center. Our calculations showed that the excitation intensity at the dipole location is enhanced 7 times when excited from top by a lens with 0.95 N.A. at 532 nm in comparison to the excitation intensity at the same dipole location in a diamond slab. The intensity profile of the resonant mode is shown in the inset of the Fig. 5(a). The resonant mode is localized closer to the aperture-air interface. We note that such a property can be useful for certain applications such as magnetometry, where NV center needs to be closer to the surface. Another important concern, in addition to the emission rate enhancement, is the collection efficiency. The photon collection efficiency from NV centers in bulk diamond slabs is rather small, Fig. 5(b). It is around 4-4.5% for the polarization that we considered. We show the collection efficiency from a dipole when it is placed inside the plasmonic aperture in Fig. 5(b). The total collection efficiency, which is defined as the ratio of the photons collected by a 0.95 N.A. lens to the total number of photons emitted, is larger than 30% over the spectrum of the NV emission. As a result, the use of plasmonic apertures significantly improves the collection efficiency, up to 10 times.

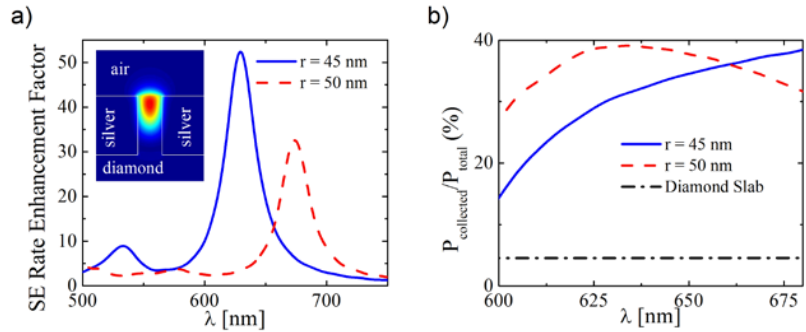

Fig. 5. (a) SE rate enhancement spectrum for r = 45 and 50 nm apertures. Inset: mode profile. (b) Total collection efficiency from a plasmonic aperture and a diamond slab when a lens with 0.95 N.A. is used.

Finally, we consider the structures shown in Figs. 1(c) and (d). The structure shown in Fig. 1(d) consists of either a single ring or multiple concentric rings around the aperture. Both structures, Figs. 1(c) and (d), are designed such that the apertures are fabricated on the bottom face of a diamond slab and the photons are collected from top. The whole bottom side of the diamond slab is covered with a thick layer of silver. These structures further facilitate fabrication and overcome issues with stability of silver in ambient conditions. The SE rate enhancement spectra for r = 55 and 60 nm apertures, Fig. 1(c), are plotted in Fig. 6(a). SE rate is enhanced up to 40 and 28 times, respectively. In this case, the resonant mode is mostly localized around the middle of the aperture (Fig. 6(a) inset). In addition, we show the relative coupling efficiencies into possible decay channels (surface plasmons, non-radiative channels, and radiation modes) in Fig. 6(b). A significant portion of the photons, up to 60%, are coupled to the surface plasmons at diamond-silver interface. On the other hand, only up to 25% of the photons are coupled to the radiation modes. Moreover, due to diamond's high refractive index, n = 2.41, only a small fraction of upward emitted photons could be collected with an objective lens. In Figure 7(a) we show the emission efficiency within the critical cone defined by the total internal reflection, in the case of structure with r = 55 nm aperture. On average, only 5% of the total emitted photons are confined within the critical cone. This collection efficiency is comparable with collection efficiency of NV in bulk, but can further be increased as we discuss next.

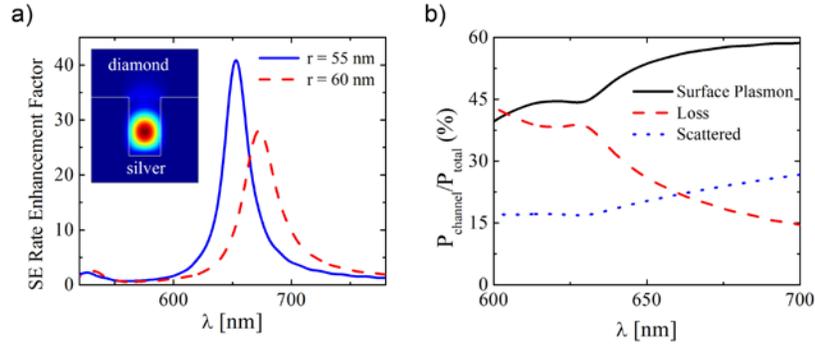

Fig. 6. (a) SE Rate enhancement spectrum of a dipole inside the plasmonic apertures. Inset: mode profile. (b) Coupling efficiency in to surface plasmons, non-radiative channels and radiation modes.

As a first improvement, we considered the structure shown in Fig. 1(d) with a single ring around the aperture. Such circular slot, when placed at well-chosen location, can scatter surface plasmon and increase the collection efficiency of the structure. Results for three different choices of Rb (280, 300, and 320 nm) are plotted in Fig. 7(a), and it can be seen that 3-fold increase in the collection efficiency can be expected, and the total of ~15% of all emitted photons can be collected with the objective lens. The maximum efficiency is obtained when the radius of the ring, Rb, is closer to the surface plasmon wavelength supported by the diamond-silver interface. Further improvement can be made by adding more rings around the aperture. This can be seen as adding second order grating that helps the extraction efficiency of surface plasmons. In fact, by just adding two more rings it is possible to increase the power emission into the critical cone up to 25%, Fig. 7(b). This high collection efficiency can be expected having in mind the far field power pattern of the structure (inset of Fig. 7(b)) that features a strong main lobe in the center, responsible for highly directional emission, and several side lobes with smaller intensity.

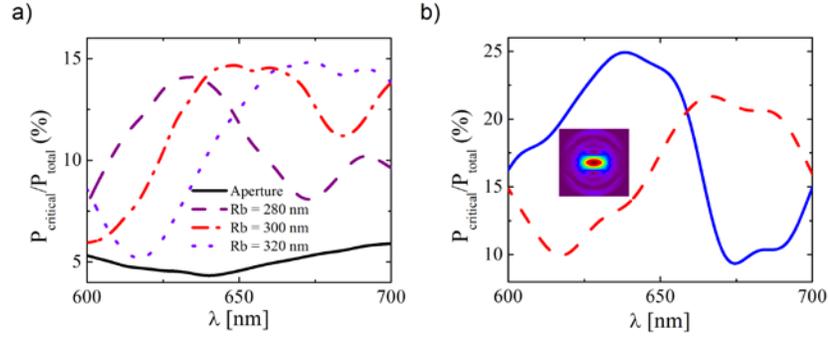

Fig. 7. (a) Emission efficiency in to the critical cone from a bare aperture and apertures surrounded by a single ring. (b) Emission efficiency in to the critical cone from apertures surrounded by 3 concentric rings. Solid blue curve: Rb = 320 nm, W = 100 nm and P = 260 nm. Red dashed curve: Rb = 320 nm, W = 100 nm and P = 275 nm.

**Conclusion**

In summary, we explored the efficiency of plasmonic aperture resonators for diamond NV center applications. We proposed three different structures for practical applications. Our results show that SE rate can be enhanced up to 50 times and collection efficiencies as high as 40% can be achieved. These plasmonic apertures can be fabricated on diamond samples by the use of existing diamond nano-fabrication methods, and eliminate the use of any complicated alignment procedure. In addition, when combined with ion-implantation, the position of the NV can be controlled deterministically along the length of the waveguide and offer maximal enhancement. This simple architecture can further be extended by coupling the plasmonic apertures to on-chip components such as plasmon or dielectric waveguides, as well as to other near-by apertures. We believe plasmonic apertures can be used to improve the efficiency of a broad range of diamond NV center applications, such as polarization entangled photon sources, magnetometers, and single photon sources.

**Acknowledgments**

This work was supported in part by NSF NIRT grant (ECCS-0708905), DARPA QuEST program, and by Sloan Foundation. Irfan Bulu acknowledges many fruitful discussions with Yinan Zhang and Emre Togan.